\documentclass[aps,prl,twocolumn,nofootinbib,groupedaddress,superscriptaddress
,longbibliography]{revtex4-2}

\usepackage{amsmath,amssymb}
\usepackage{multirow}
\usepackage{bm}
\usepackage{mathtools}
\usepackage{dsfont}
\usepackage{amsfonts}
\usepackage[normalem]{ulem}
\usepackage{url}
\usepackage{wasysym}
\usepackage{bbm}
\usepackage{hyperref}

\newcommand{\R}[1]{\textcolor{red}}
\newcommand{\B}[1]{\textcolor{blue}}
\newcommand{\G}[1]{\textcolor{green}}
\begin{document}
\title{Clifford Circuits Augmented Time-Dependent Variational Principle }

\author{Xiangjian Qian}
\affiliation{Key Laboratory of Artificial Structures and Quantum Control (Ministry of Education),  School of Physics and Astronomy, Shanghai Jiao Tong University, Shanghai 200240, China}

\author{Jiale Huang}
\affiliation{Key Laboratory of Artificial Structures and Quantum Control (Ministry of Education),  School of Physics and Astronomy, Shanghai Jiao Tong University, Shanghai 200240, China}

\author{Mingpu Qin} \thanks{qinmingpu@sjtu.edu.cn}
\affiliation{Key Laboratory of Artificial Structures and Quantum Control (Ministry of Education),  School of Physics and Astronomy, Shanghai Jiao Tong University, Shanghai 200240, China}

\affiliation{Hefei National Laboratory, Hefei 230088, China}

\date{\today}


\begin{abstract}
The recently proposed Clifford Circuits Augmented Matrix Product States (CA-MPS) (arXiv:2405.09217) seamlessly augments Density Matrix Renormalization Group with Clifford circuits. In CA-MPS, the entanglement from stabilizers is transferred to the Clifford circuits which can be easily handled according to the Gottesman-Knill theorem. As a result, MPS needs only to deal with the non-stabilizer entanglement, which largely reduce the bond dimension and the resource required for the accurate simulation of many-body systems. In this work, we generalize CA-MPS to the framework of Time-Dependent Variational Principle (TDVP) for time evolution simulations. In this method, we apply Clifford circuits to the resulting MPS in each TDVP step with a two-site sweeping process similar as in DMRG, aiming at reducing the entanglement entropy in the MPS, and the Hamiltonian is transformed accordingly using the chosen Clifford circuits. Similar as in CA-MPS, the Clifford circuits doesn't increase the number of terms in the Hamiltonian which makes the overhead very small in the new method. We test this method in both XXZ chain and two dimensional Heisenberg model. The results show that the Clifford circuits augmented TDVP method can reduce the entanglement entropy in the time evolution process and hence makes the simulation reliable for longer time. The Clifford circuits augmented Time-Dependent Variational Principle provides a useful tool for the simulation of time evolution process of many-body systems in the future. 
\end{abstract}

\maketitle
{\em Introduction --}
Solving strongly correlated quantum many-body systems is one of the major challenges in modern physics, due to the exponential growth of the dimension of the underlying Hilbert space and the intricate quantum correlations involved. To deal with these challenges, powerful numerical methods were developed in the last decades \cite{PhysRevX.5.041041}. Density Matrix Renormalization Group (DMRG) \cite{PhysRevLett.69.2863} and Matrix Product States (MPS) \cite{RevModPhys.77.259,1992CMaPh.144..443F,10.5555/2011832.2011833,SCHOLLWOCK201196,RevModPhys.93.045003} provide a useful numerical framework for analyzing and simulating one-dimensional (1D) quantum many-body systems.

\begin{figure*}[t]
    \includegraphics[width=170mm]{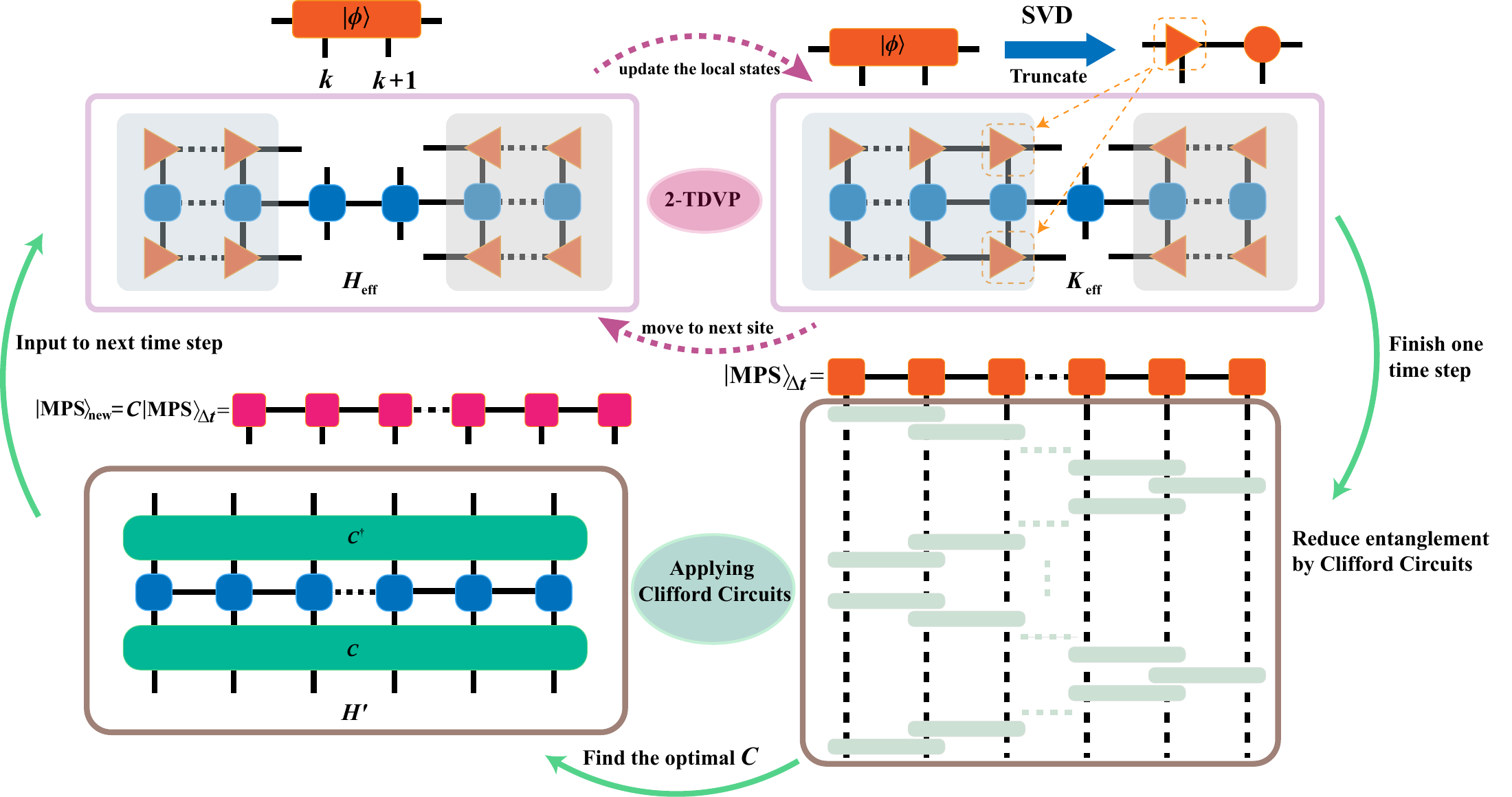}
       \caption{The CA-TDVP method is divided into three successive steps: 1. Evolve the initial state $|\text{MPS}\rangle$ according to the Hamiltonian $H$ by a time step $\Delta t$ to obtain $|\text{MPS}\rangle_{\Delta t} $ using the two-site TDVP method. 2. Apply Clifford circuits $\mathcal{C}$ to $|\text{MPS}\rangle_{\Delta t}$ to reduce its entanglement, resulting in a new state $|\text{MPS}\rangle_{\text{new}} \approx \mathcal{C} |\text{MPS}\rangle_{\Delta t}$. In this step, a two-site DMRG like sweeping process is performed. 3. Perform a transformation on the original Hamiltonian to obtain $H^{\prime}= \mathcal{C} H \mathcal{C}^\dagger $. Proceed the time evolution using the transformed Hamiltonian $H^{\prime}$, starting from the new initial state $|\text{MPS}\rangle_{\text{new}}$.}
       \label{CA-MPS}
\end{figure*}


But the entanglement entropy encoded in MPS is bounded by $\log(D)$ ($D$ is the bond dimension) which hampers the application of MPS to large two-dimensional (2D) systems, even though DMRG can now handle quite wider cylinders for spin systems \cite{PhysRevB.109.L161103, huang2024valencebondsolidstate} by pushing the bond dimension to very large numbers, thanks to the improvement of algorithm efficiency and the increase of computational power. Recently, Clifford Circuits Augmented Matrix Product States (CA-MPS) \cite{qian2024augmentingdensitymatrixrenormalization} method was proposed to enhance the power of MPS in the simulation of many-body systems. The idea of augmenting MPS with other ingredients to enlarge the encoded entanglement was proposed in the development of Fully-augmented Matrix Product States (FA-MPS) \cite{Qian_2023}, where general disentanglers (unitary transformation) are used to reduce the entanglement in the ground state of the studied system. In CA-MPS, MPS are augmented by a special group of disentanglers, i.e., Clifford circuits \cite{Nielsen_Chuang_2010}.    

Clifford circuits are composed exclusively of Clifford gates (Hadamard, S, and Controlled-NOT gates) \cite{Nielsen_Chuang_2010}, which can be efficiently simulated on a classical computer according to the Gottesman-Knill theorem \cite{gottesman1997stabilizer,PhysRevA.70.052328,PhysRevA.73.022334}. The states that can be prepared under these constraints are known as stabilizer states \cite{lami2024learning,PhysRevA.70.052328,PhysRevA.73.022334,sun2024stabilizer,gottesman1997stabilizer,PhysRevLett.128.050402}, which can manifest significant entanglement yet remain simulatable. the Gottesman-Knill theorem serves as a compelling example, emphasizing that while entanglement is a vital quantum resource, its presence alone does not necessarily make a computational problem classically hard. The contribution of entanglement which can't be captured by Clifford circuits are known as non-stabilizerness or “magic” \cite{PhysRevLett.128.050402,lami2024learning,tarabunga2024nonstabilizerness,PhysRevLett.131.180401,frau2024nonstabilizerness,PhysRevLett.112.240501,PRXQuantum.3.020333,PhysRevA.71.022316,10.21468/SciPostPhys.16.2.043,masotllima2024stabilizertensornetworksuniversal}, which is the key for universal quantum computing.

Other than ground state simulation, the time evolution process of many-body systems provides more information about the studied system. But the time evolution simulation using MPS or other tensor network based methods is more challenging because the entanglement entropy usually increases linearly with the evolution time which means the required bond dimension for MPS increases exponentially even for 1D systems \cite{PAECKEL2019167998}. In this work, we attempt to generalize CA-MPS to the simulation of time evolution process of quantum many-body systems. 

In the framework of Tensor Networks, time-evolving block decimation (TEBD) \cite{PhysRevLett.93.040502,Daley_2004} and the Time-Dependent Variational Principle (TDVP) \cite{PhysRevLett.107.070601,PhysRevB.94.165116} are two common choices for time evolution simulations. We find that directly augmenting TEBD with Clifford circuits breaks the locality of the projection operators which prevents a simple implementation of the algorithm. But TDVP shares many similarities with DMRG and the augmenting of TDVP with Clifford circuits can be implemented quite straightforwardly, as we did in CA-MPS \cite{qian2024augmentingdensitymatrixrenormalization}.

Another question we also want to answer in this work is whether the major contribution of entanglement growth in the time evolution of typical many-body systems comes from stabilizers or non-stabilizerness. If stabilizers dominate the entanglement growth, we anticipate that augmenting TDVP with Clifford circuits will be extremely useful and this new method can enable us to accurately simulate long-time evolution.

In the rest of this work, we discuss the details of the Clifford Circuits Augmented Time-Dependent Variational Principle (CA-TDVP) method. We also test the performance of CA-TDVP in both 1D and 2D systems. Our results seem to indicate that in the long-time evolution, the increase of entanglement entropy mainly comes from the contribution of non-stabilizerness. Nevertheless, we find that CA-TDVP can enable the simulation of evolution for a longer time than the TDVP with the same bond dimension. 


{\em Clifford Circuits Augmented Time-Dependent Variational Principle (CA-TDVP)--}
The most straightforward approach for simulating the time evolution of an initial state in the form of tensor network states, such as MPS, is through the Trotter decomposition of the time evolution operator $e^{-iH\Delta t}$, which transforms the Hamiltonian dynamics into a discrete quantum circuit composed of unitary gates acting locally between neighboring lattice sites in the system, making it particularly suitable for simulation within tensor network state framework. The well-known TEBD \cite{PhysRevLett.93.040502,Daley_2004} method belongs to this class, which efficiently manages real-time evolution by decomposing the global evolution into a sequence of local unitary gates.

However, directly augmenting TEBD with Clifford circuits breaks the locality of the projection operators, i.e., long-range interactions appear as in CA-MPS \cite{qian2024augmentingdensitymatrixrenormalization}, which prevents a simple implementation of the algorithm. But another time evolution method, TDVP \cite{PhysRevLett.107.070601,PhysRevB.94.165116}, has proven to be a successful approach for handling long-range interactions. TDVP also shares many similarities with DMRG which makes the augmenting of TDVP with Clifford circuits straightforward. The implementation is quite similar to what we did previously in changing a DMRG code to implement CA-MPS. 

\begin{figure*}[t]
    \includegraphics[width=75mm]{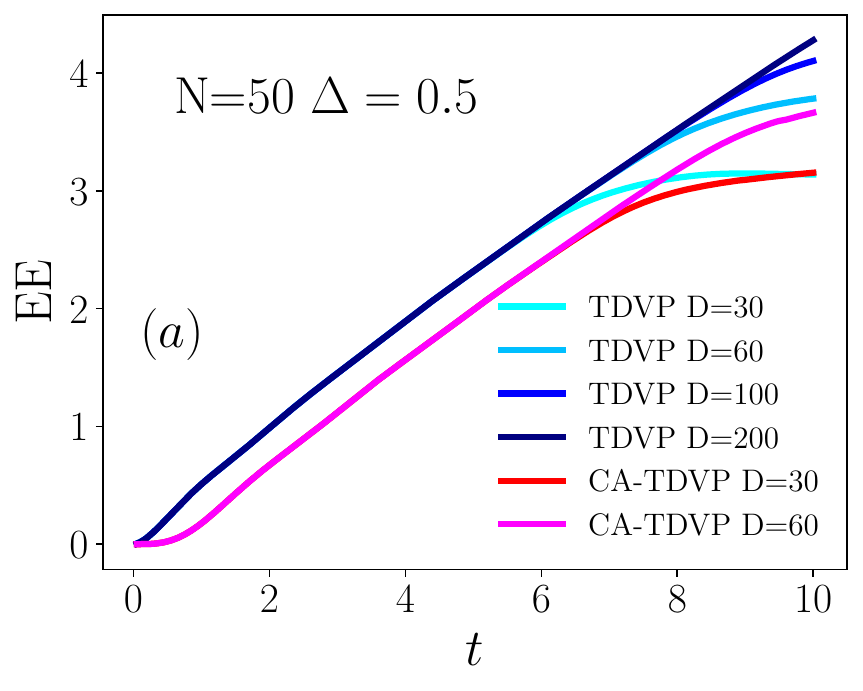}
    \includegraphics[width=82.5mm]{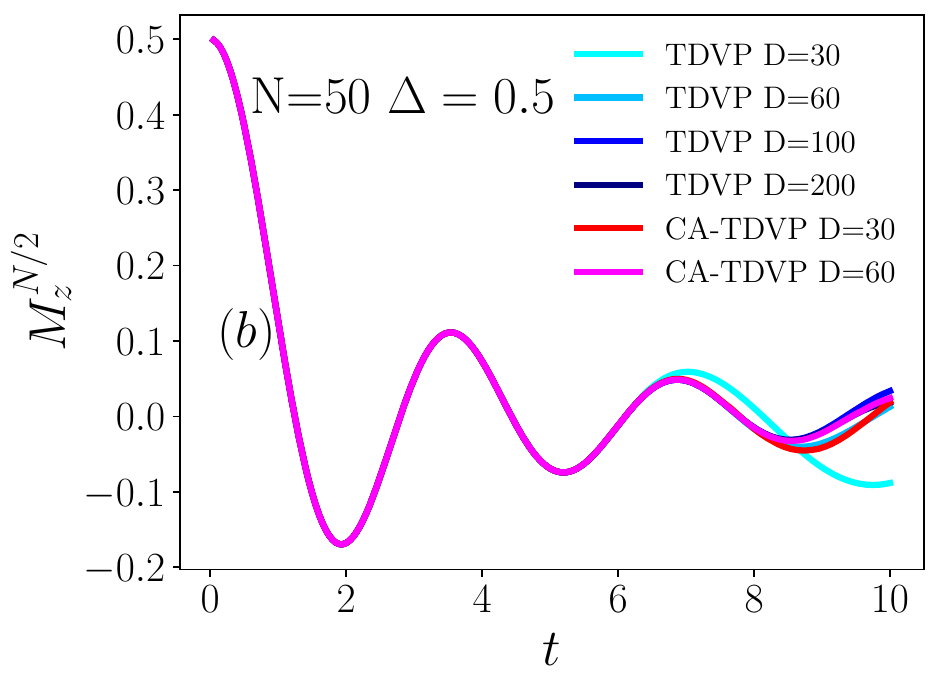}
       \caption{
       The CA-TDVP results for the 1D XXZ model. The Hamiltonian is shown in Eq.~(\ref{XXZ_H}). The length of the chain is $N=50$ and $\Delta=0.5$. The time step in TDVP is set as $\Delta t=0.05$. The initial state is set as the N\'eel state $|\uparrow \downarrow \uparrow \downarrow \cdots \uparrow \downarrow \uparrow \downarrow\rangle$. Open boundary conditions are considered. (a) The half-chain entanglement entropy (EE) as a function of time $t$. (b) The measured $M_z^{N/2}$ at the center site as a function of time $t$. From (a), we can find the reduction of entanglement entropy with Clifford circuits in CA-TDVP, compared to the TDVP results. For short-time CA-TDVP sweeps, the entanglement entropy is nearly reduced to zero, indicating that the short-time EE is primarily contributed by stabilizer. But the results also show the major contribution of entanglement growth in long-time evolution comes from the non-stabilizerness. In (b), we can see that with the same bond dimension $D$, CA-TDVP can support longer time evolution than TDVP.} 
       \label{XXZ}
\end{figure*}

The two-site TDVP method decomposes \cite{PhysRevLett.107.070601,PhysRevB.94.165116} the short-time evolution $e^{-iH\Delta t}$  of an MPS into a sweep process analogous to the DMRG algorithm. Specifically, in a left-to-right sweep, one deals with the effective Hamiltonians $H_{\text{eff}}=\sum_{i=1}^{m} a_i A_{i,k-1} \otimes \sigma_{i,k} \otimes \sigma_{i,k+1} \otimes B_{i,k+2}$ and $K_{\text{eff}}=\sum_{i=1}^{m} a_i A_{i,k} \otimes \sigma_{i,k+1} \otimes B_{i,k+2}$.  Here, the Hamiltonian has been rewritten as a sum of Pauli strings $P=\sigma_1 \otimes \sigma_2\cdots \otimes \sigma_N$ ($\sigma_i\in \{I, \sigma^x, \sigma^y, \sigma^z\}$): $H=\sum_{i=1}^{m}a_i P_i$, $A_{i,k}, B_{i,k}$ are the so-called left and right environment for $P_i$ at site $k$, $\sigma_{i,k}$ is the Pauli matrix of $P_i$ at site $k$ and $a_i$ is the associated interaction strength of $P_i$ \cite{RevModPhys.77.259}. In two-site TDVP, we first update the local states $|\phi\rangle$ associated with local tensors $M_k,M_{k+1}$ according to $H_{\text{eff}}$: $|\phi\rangle \coloneq e^{-iH_{\text{eff}}\Delta t}|\phi\rangle$. Then, we perform a Singular Value Decomposition (SVD) on $|\phi\rangle$ to obtain the updated tensors $M_k,M_{\text{k+1}}$, as well as the updated left environment $A_{i,k}$. Finally, we update the local state $|\psi\rangle$ associated with local tensor $M_{k+1}$ according to $K_{\text{eff}}$: $|\psi\rangle \coloneq e^{iK_{\text{eff}}\Delta t}|\psi\rangle$. Sweeping from $k=1$ to $k=N$ ($N$ is the system size) evolves the system for a short time $\Delta t$ with a small error: $|\text{MPS} \rangle_{\Delta t} \approx e^{-iH\Delta t} |\text{MPS} \rangle$.

Because the entanglement entropy in $e^{-iH t} |\text{MPS}\rangle$ usually increases linearly with time $t$, the TDVP procedure described above can't simulate long-time evolution with a given bond dimension $D$, resulting in a significant truncate error $\epsilon$ after several sweeps. 

We can incorporate Clifford circuits to mitigate the entanglement \cite{PhysRevLett.112.240501, Shaffer_2014} in $|\text{MPS}\rangle$, as demonstrated in CA-MPS \cite{qian2024augmentingdensitymatrixrenormalization}, thereby enabling longer time evolution simulations. Specifically, by applying a Clifford circuit $\mathcal{C}$, we obtain a new MPS state $|\text{MPS}\rangle_{\text{new}} = \mathcal{C} |\text{MPS}\rangle$, where $|\text{MPS}\rangle_{\text{new}}$ is less entangled than the original $|\text{MPS}\rangle$. To ensure that the physical observables remain unchanged, we need to introduce a corresponding transformation of the Hamiltonian, $H^{\prime}=\mathcal{C}H\mathcal{C}^{\dagger}$ which can be efficiently performed using the stabilizer tableau formalism \cite{gottesman1997stabilizer,PhysRevA.70.052328,PhysRevA.73.022334} (same transformation is also needed for physical observable operators). This can be understood from the fact that $\langle\text{MPS}|H|\text{MPS}\rangle=\langle\text{MPS}|\mathcal{C}^{\dagger}\mathcal{C}H\mathcal{C}^{\dagger}\mathcal{C}|\text{MPS}\rangle=\sideset{_{\text{new}}}{}{\mathop{\langle\text{MPS}}}|H^{\prime}|\text{MPS}\rangle_{\text{new}}$. This transformation ensures that the expectation values of the Hamiltonian are preserved in the new, less entangled MPS state $|\text{MPS}\rangle_{\text{new}}$.

Thus, the CA-TDVP method is performed as follows:
\begin{enumerate}
\item Evolve the initial state $|\text{MPS}\rangle$ according to the Hamiltonian $H$ by a time step $\Delta t$ to obtain $|\text{MPS}\rangle_{\Delta t} $ using the two-site TDVP method.
\item Apply Clifford circuits $\mathcal{C}$ to $|\text{MPS}\rangle_{\Delta t}$ to reduce its entanglement, resulting in a new state $|\text{MPS}\rangle_{\text{new}} = \mathcal{C} |\text{MPS}\rangle_{\Delta t}$.
\item Perform a transformation on the original Hamiltonian to obtain $H^{\prime}= \mathcal{C} H \mathcal{C}^\dagger $. Crucially, any observables we wish to measure must also undergo this transformation.
\item Repeat step 1. Proceed the time evolution using the transformed Hamiltonian $H^{\prime}$, starting from the new initial state $|\text{MPS}\rangle_{\text{new}}$.
\end{enumerate}  
An illustration of this process is shown in Fig.~\ref{CA-MPS}. In step 2, we perform a two-site sweeping to choose the two-qubit Clifford circuits which minimize the entanglement entropy, from the $720$ possible gates \cite{10.1063/1.4903507,PhysRevB.100.134306,PhysRevA.87.030301}, similarly as we did in \cite{qian2024augmentingdensitymatrixrenormalization}. We also tried to incorporate Clifford circuits in the SVD step in TDVP, but the results are not as good as the scheme we described above. The reason could be that the SVD step in TDVP is followed by a reversal evolution step, which is different from the SVD step in DMRG.

\begin{figure*}[t]
    \includegraphics[width=74mm]{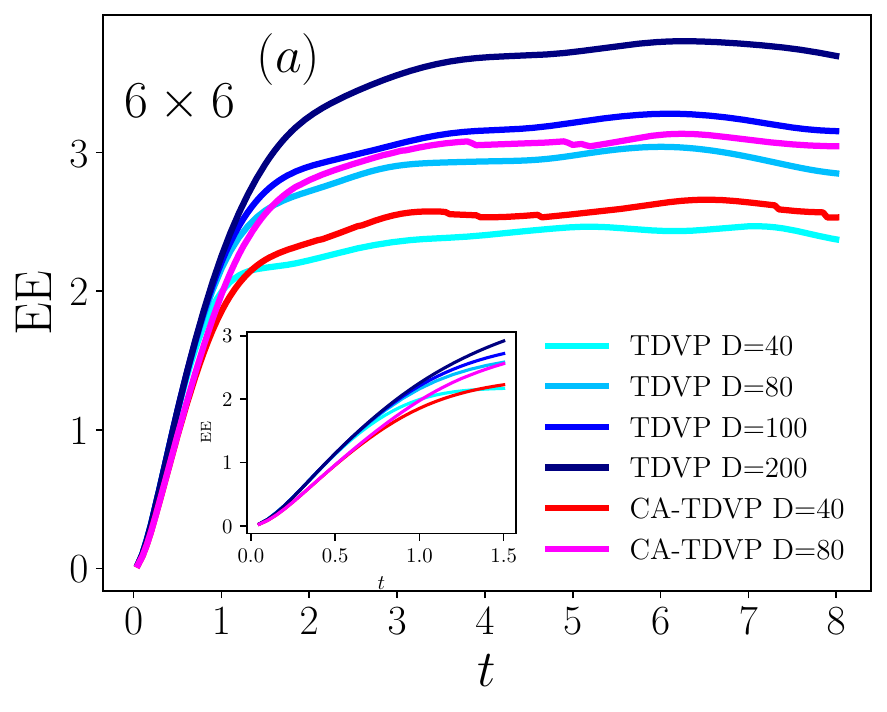}
    \includegraphics[width=80mm]{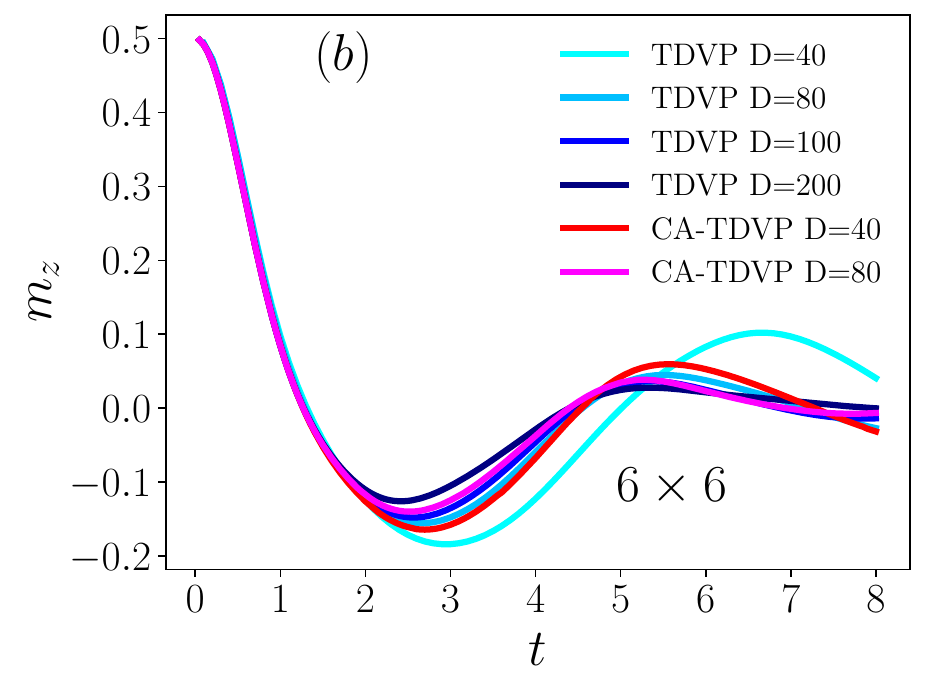}
       \caption{
       The CA-TDVP results for the 2D Heisenberg model. The system is with size $6\times 6$ under open boundary conditions. The initial state is set as a 2D N\'eel state $|\uparrow \downarrow \cdots \uparrow \downarrow, \downarrow \uparrow \cdots \downarrow \uparrow,\cdots, \uparrow \downarrow \cdots \uparrow \downarrow, \downarrow \uparrow \cdots \downarrow \uparrow\rangle$. (a) The entanglement entropy (EE) at the center bond in the MPS part as a function of time $t$. The inset shows the short-time behavior of EE.  (b) The measured N\'eel order parameter $m_z=1/N \sum_{i,j} (-1)^{i+j}\langle S^z_{i,j}\rangle$ as a function of time $t$. The 2D Heisenberg results are similar to the results for the XXZ chain in Fig.~\ref{XXZ} but with a faster increase of entanglement entropy with time.}
       \label{2D_Hei}
\end{figure*}

{\em Results for XXZ chain --}
We first test the CA-TDVP method on the 1D XXZ model under open boundary conditions (OBC). The Hamiltonian of the model is defined as
\begin{equation}
	H = \sum_i^N S^x_{i} S^x_{i+1}+S^y_{i} S^y_{i+1}+ \Delta S^z_{i} S^z_{i+1}
 \label{XXZ_H}
\end{equation}
where $S^x_i, S^y_i$ and $S^z_i$ are the spin-1/2 operator on site $i$.

Fig.~\ref{XXZ} shows the results for a chain with length $N=50$ and $\Delta=0.5$. The time step in TDVP is set as $\Delta t=0.05$. The initialed state is the N\'eel state $|\uparrow \downarrow \uparrow \downarrow \cdots \uparrow \downarrow \uparrow \downarrow\rangle$. The local magnetization ($M_z^{N/2}$) at the center site and the half-chain entanglement entropy (EE) as a function of time $t$ are plotted. From Fig.~\ref{XXZ} (a), we can find the reduction of entanglement entropy with Clifford circuits in CA-TDVP, compared to the TDVP results. For short-time CA-TDVP sweeps, the entanglement entropy is nearly reduced to zero, indicating that the short-time EE is primarily contributed by stabilizer. But the results also show the major contribution of entanglement growth in long-time evolution comes from the non-stabilizerness. In (b), we can see that with the same bond dimension $D$, CA-TDVP can support longer time evolution than TDVP. To achieve the same accuracy, CA-TDVP only requires a bond dimension that is half or third of the bond dimension of TDVP. 


{\em Results for 2D Heisenberg model --}
We also test the CA-TDVP method on the 2D Heisenberg model, with Hamiltonian:
\begin{equation}
    H = J\sum_{\langle i,j \rangle}S_{i} \cdot S_{j}
\end{equation}
where $S_i$ is the spin-1/2 operator on site $i$, and the summations are taken over nearest-neighbor ($\langle i,j \rangle$) pairs. Again, the initial state is set as a 2D N\'eel state $|\uparrow \downarrow \cdots \uparrow \downarrow, \downarrow \uparrow \cdots \downarrow \uparrow,\cdots, \uparrow \downarrow \cdots \uparrow \downarrow, \downarrow \uparrow \cdots \downarrow \uparrow\rangle$. The TDVP time step is set as $\Delta t=0.05$.
Here, we consider a $6\times 6$ lattice with open boundary conditions. Fig.~\ref{2D_Hei} shows the measured N\'eel order parameter $m_z=1/N \sum_{i,j} (-1)^{i+j}\langle S^z_{i,j}\rangle$ and the entanglement entropy at the center bond as a function of evolution time $t$. The 2D Heisenberg results are similar to the results for the XXZ chain in Fig.~\ref{XXZ} but with a faster increase of entanglement entropy with time. The CA-TDVP is again able to achieve the same accuracy as TDVP with about half the bond dimension. For example, the CA-TDVP results with $D=40$ are comparable to those of TDVP with $D=80$ as shown in Fig.~\ref{2D_Hei} (b). Interestingly, the increasing rate of the entanglement entropy over time $t$ in CA-TDVP is smaller than TDVP, which is contrary to the 1D part, where both CA-TDVP and TDVP's results have the same increasing rate.

{\em Discussion --}
From the CA-TDVP results of the 1D XXZ model and 2D Heisenberg model, we find that the reduced entanglement is not as significant as in the ground state simulation \cite{qian2024augmentingdensitymatrixrenormalization}, which suggests that the increase of entanglement entropy in time evolution is mainly contributed by non-stabilizerness. However, the results of magnetization indicate that CA-TDVP allows for much longer simulation time compared to pure TDVP calculations. Since the reduction in entanglement is more pronounced for ground state simulations, one could use CA-TDVP to explore the quench dynamics of a quantum system \cite{2017NatPh..13..246K,PhysRevX.3.031015}. This involves suddenly changing the Hamiltonian and then studying the resulting dynamics. In such scenarios, the initial state is typically not a product state as we study in this work and CA-TDVP can offer substantial benefits during the initial TDVP sweeps, enabling even longer time simulations. One could also incorporate a small number of T gates \cite{Leone2024learningtdoped, PhysRevA.71.022316} into CA-TDVP, enabling a bounded increase of the number of terms in $H^{\prime}$, further enhancing the performance of CA-TDVP.

{\em Conclusion and Perspective --}
In this study, we generalize CA-MPS to CA-TDVP to investigate the time evolution process of quantum many-body systems. In CA-TDVP, Clifford circuits are applied to the resulting MPS in TDVP to reduce the entanglement entropy to enable longer time evolution. Our numerical tests on both the XXZ chain and the 2D Heisenberg mode shows the effectiveness of CA-TDVP. Even though the results indicate non-stabilizerness dominates the entanglement increase in time evolution, CA-TDVP can reduce the bond dimension by almost a half to achieve the same accuracy as pure TDVP. An interesting relaved direction is the application of CA-TDVP to the finite temperature simulation, where imaginary instead of real-time evolution is performed \cite{PhysRevLett.130.226502, PhysRevResearch.3.L032017,PhysRevB.104.214302}. We are now in the process of testing the performance of CA-TDVP in finite temperature simulations.  


\textbf{Note added:}
We thank the authors of \cite{Antonio0701} for sharing with us the preprint of \cite{Antonio0701} before it was posted on arXiv. In \cite{Antonio0701}, a similar idea of augmenting TDVP with Clifford circuits is also proposed. A minor difference is that we use the 2-site TDVP while 1-site TDVP was adopted in \cite{Antonio0701}. 

\begin{acknowledgments}
\textbf{Acknowledgments:} The calculation in this work is carried out with TensorKit \cite{foot7}. The computation in this paper were run on the Siyuan-1 cluster supported by the Center for High Performance Computing at Shanghai Jiao Tong University. MQ acknowledges the support from the National Natural Science Foundation of China (Grant No. 12274290), the Innovation Program for Quantum Science and Technology (2021ZD0301902), and the sponsorship from Yangyang Development Fund.
\end{acknowledgments}

\bibliography{main}

\end{document}